\documentclass{caosp305}

\usepackage{graphicx}

\usepackage{natbib}
\articleNo{1}
\pubyear{2017}
\volume{00}
\volnumber{0}
\firstpage{1}
\received{October 31, 2017}
\accepted{October 31, 2017}

\def\BibTeX{{\rm B\kern-.05em{\sc i\kern-.025em b}\kern-.08em
             T\kern-.1667em\lower.7ex\hbox{E}\kern-.125emX}}

\begin{document}

%
\hauthor{S.\,Vennes, A.\,Kawka, L.\,Ferrario, and E.\,Paunzen}

\title{Observing and modelling magnetic fields in white dwarfs}


%
\author{
        S.\,Vennes \inst{1} 
      \and 
        A.\,Kawka \inst{1}
       \and
        L.\,Ferrario \inst{2}   
      \and 
        E.\,Paunzen \inst{3}
       }

%
\institute{
           \ondrejov, \email{vennes@asu.cas.cz}
         \and 
           Australian National University, Canberra,
           Australia
         \and 
           Masaryk University, Brno, Czech Republic
          }

\date{October 31, 2017}

\maketitle

\begin{abstract}
Our ongoing spectroscopic survey of high proper motion stars is a rich source of new magnetic white dwarfs. We present a few examples among cool white dwarfs showing the effect of field strength and geometry on the observed optical spectrum. Modelling of hydrogen and heavy element spectral lines reveals a range of uniform or markedly offset dipole fields in these objects. 
\keywords{white dwarfs -- magnetic fields -- spectra}
\end{abstract}

%
\section{Observations}

Magnetic white dwarfs account for a substantial fraction of the population of white dwarf stars \citep{2007ApJ...654..499K}. Spectroscopic surveys \citep{2012MNRAS.425.1394K} routinely uncover new candidates showing a great diversity in field strength and geometry \citep{2017A&A...607A..92L}. Our most recent observations were obtained with ESO's FOcal Reducer and low-dispersion Spectrograph 2 (FORS2) and the intermediate-dispersion X-shooter spectrograph both on ESO's Very Large Telescopes (VLTs). Detailed modelling of spectroscopic time series often reveals complex surface field structures or the presence of a close degenerate companion, as observed in the case of NLTT~12758 \citep{2017MNRAS.466.1127K}.

\section{Modeling and analysis}

We followed a methodology described in \citet{1984MNRAS.206..407M} and \citet{1989ApJ...346..444A} and modelled the field distribution in magnetic hydrogen-rich white dwarfs, known as DAH white dwarfs, using a dipole of strength $B_p$ which may be offset along the polar axis by a fraction of the radius $a_z$ and inclined with respect to the viewer at an angle $i$. We divided the surface into 450 elements along the surface longitude and latitude and integrated the emergent intensity spectrum. These model spectra describe average surface field properties at a particular time and do not account for possible blurring caused by a short rotation period.

\subsection{Hydrogen Balmer lines}

\begin{figure*}[t!]
\vspace{-0.3cm}
\includegraphics[width=0.49\textwidth,clip=25 317 548 674]{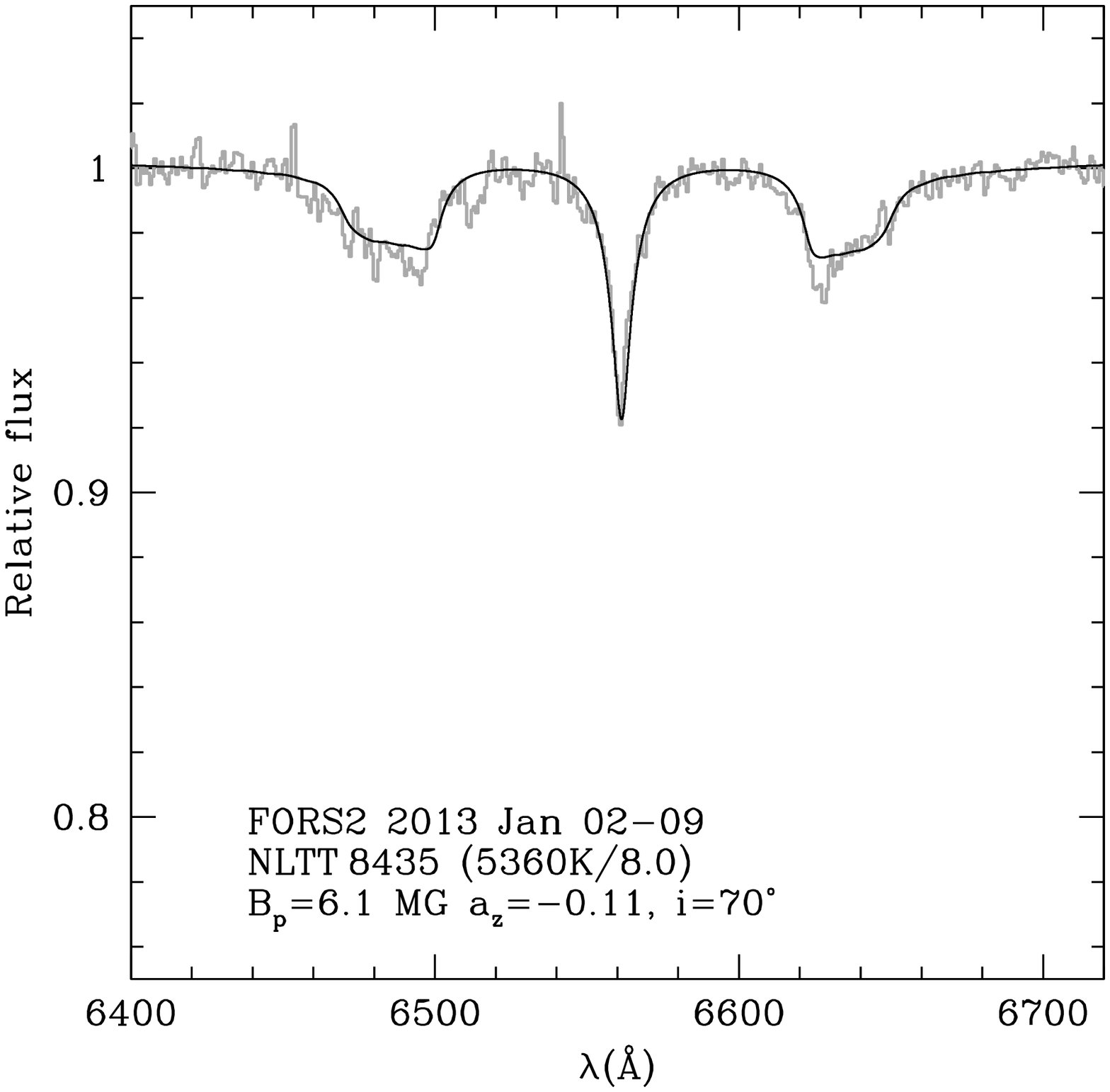}%
\includegraphics[width=0.49\textwidth,clip=25 317 548 674]{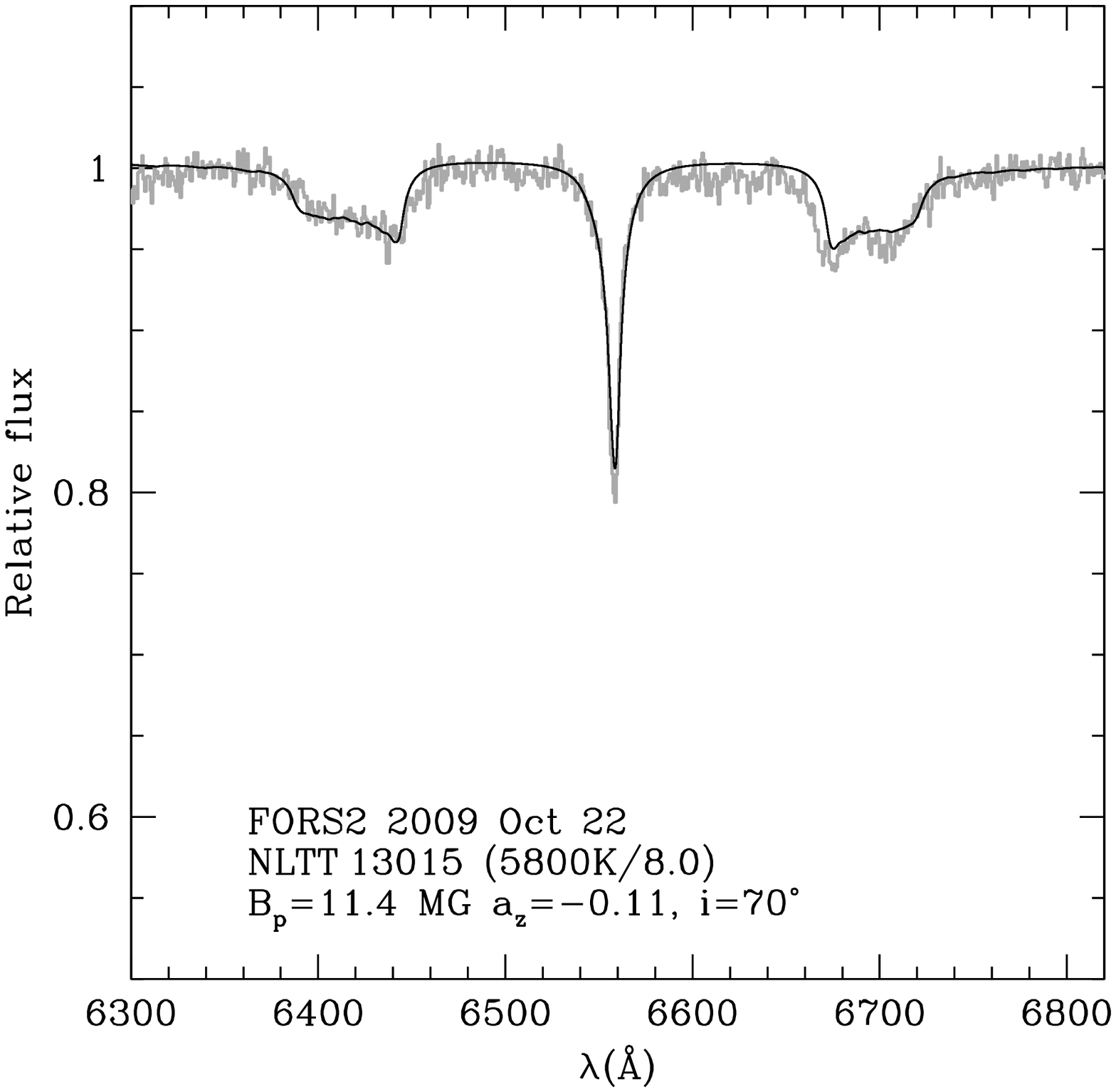}
\vspace{-0.2cm}
\caption{Observation and modelling (H$\alpha$) of the DAH white dwarfs  NLTT\,8435 (left) and NLTT\,13015 (right).}
\label{fig1}
\end{figure*}

\begin{figure*}[t!]
\vspace{-0.2cm}
\begin{center}
\includegraphics[width=0.6\textwidth,clip=25 317 548 674]{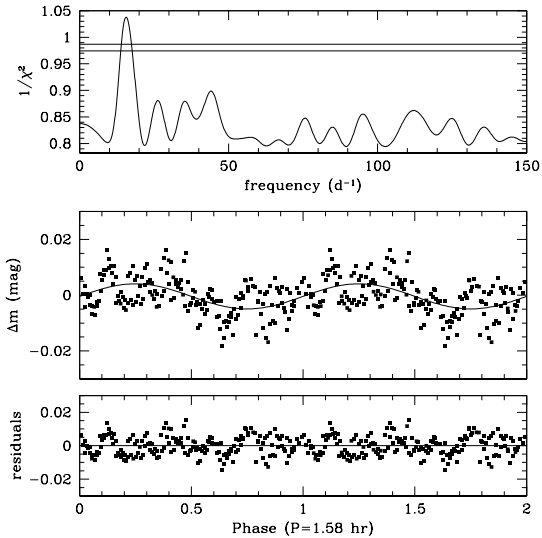}
\vspace{-0.0cm}
\caption{Photometric (R-band) time series (middle panel) and residuals (bottom panel)  of NLTT~8435 obtained with the Danish 1.54-m telescope. The period analysis finds a significant periodicity near 95 minutes (top panel).}
\label{fig2}
\end{center}
\end{figure*}

The hydrogen Balmer spectra were computed using line strengths and Zeeman shifts from \citet{1974Ap&SS..31..103G}.
The following examples illustrate the method. The new magnetic white dwarf NLTT~8435 ($B_p=6.1$\,MG) is  relatively cool ($\approx 5360$~K) and hydrogen-rich (Fig.~\ref{fig1}). Photometric time series obtained with the Danish 1.54-m telescope revealed a likely rotation period of 95 minutes (Fig~\ref{fig2}). We also observed radial velocity variations of at least 60~km\,s$^{-1}$ that are not related to surface field variations but, instead, caused by the presence of a close, unseen companion. The cool magnetic white dwarf NLTT~13015 is also hydrogen-rich and exhibits marked field variations around a mean polar field of $\approx$12~MG. Fig.~\ref{fig1} shows one of the three individual exposures obtained with FORS2: The best-fitting model implies a field strength of 11.4~MG and a small offset along the polar axis of -11\%.

\subsection{Heavy elements}

White dwarf atmospheres are often contaminated with trace heavy elements \citep{2003ApJ...596..477Z}. Some cool and polluted hydrogen-rich white dwarfs known as DAZH white dwarfs such as NLTT~7547 (Kawka et al. 2018, in preparation) and NLTT~53908 \citep{2014MNRAS.439L..90K} show strong CaH\&K lines imbedded in a magnetic field with strengths ranging from $\approx10^5$ to $10^6$ G. Other trace elements are also seen in the spectra of these objects (e.g., sodium, magnesium, aluminum, and iron) and modelling of spectral line shapes should provide additional constraints on the strength and structure of the magnetic field.

\begin{figure*}[t!]
\vspace{-0.3cm}
\includegraphics[width=0.49\textwidth,clip=25 317 548 674]{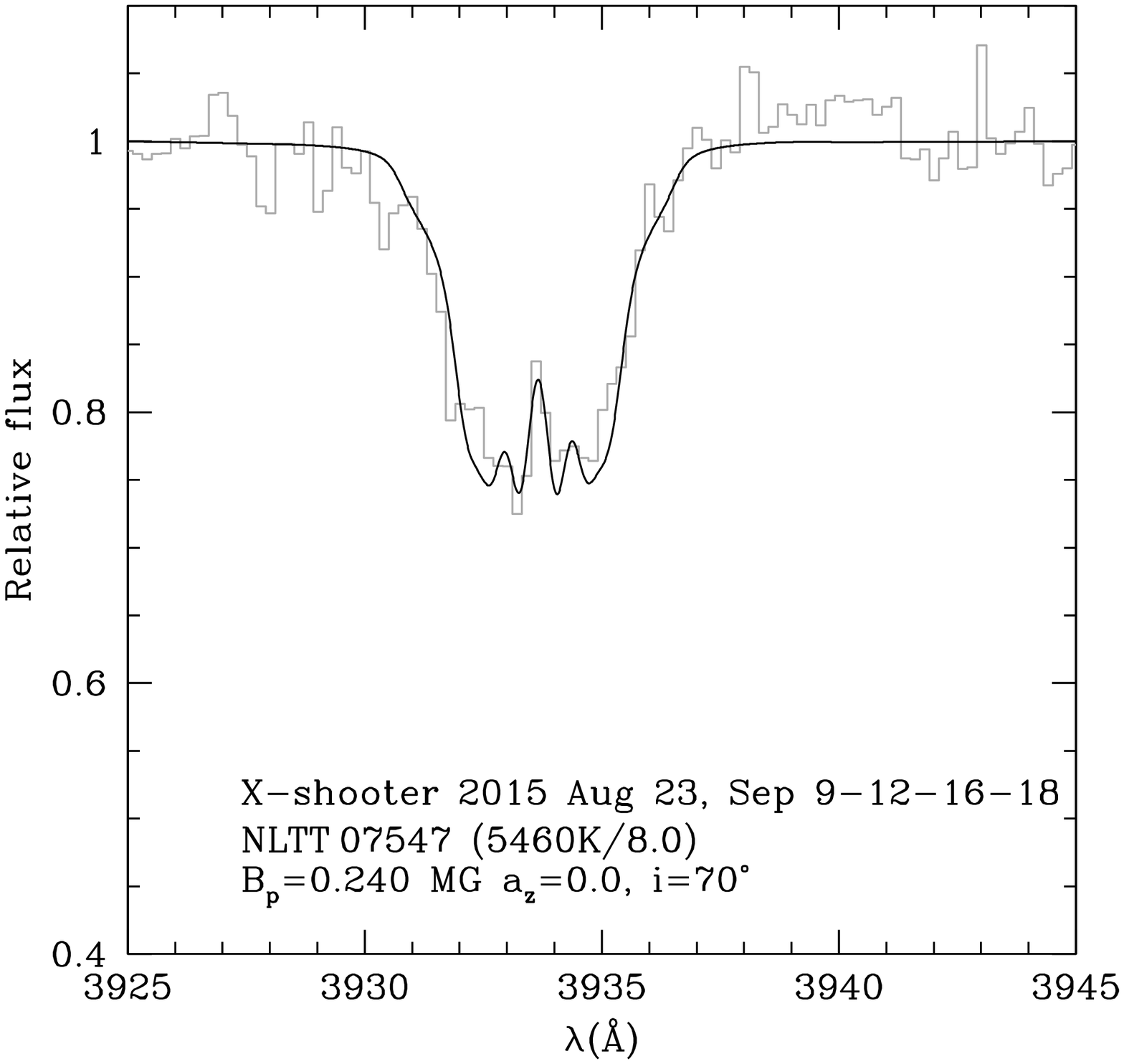}%
\includegraphics[width=0.49\textwidth,clip=25 317 548 674]{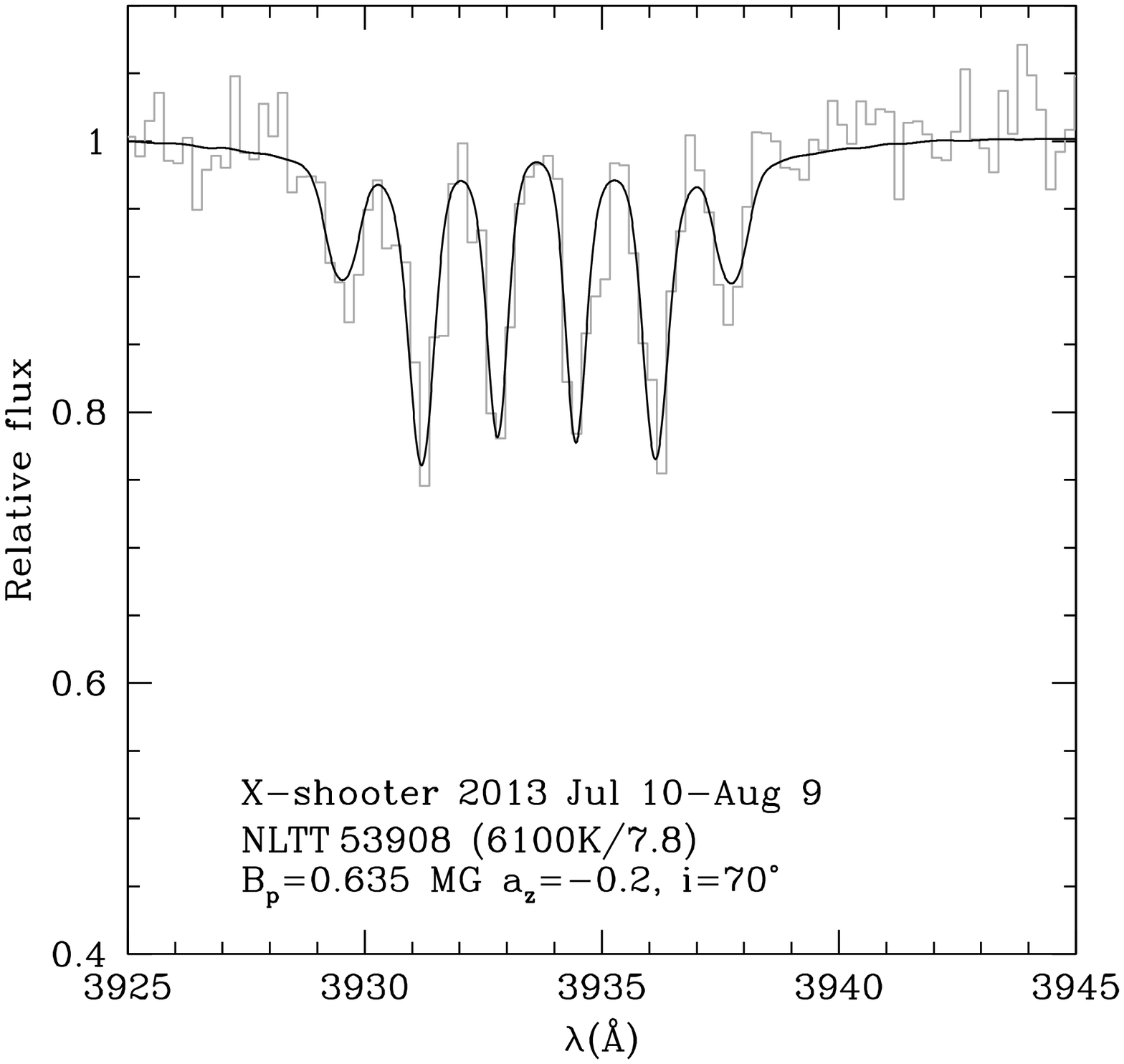}
\vspace{-0.2cm}
\caption{Observation and modelling (Ca~K) of the DAZH white dwarfs NLTT\,7547 (left) and NLTT\,53908 (right).}
\label{fig3}
\end{figure*}

We computed detailed line profiles following the procedure described in \citet{2011A&A...532A...7K} but updated with offset dipole field distributions described above and assuming quadratic Zeeman line splitting following \citet{2004ASSL..307.....L}. The updated Zeeman patterns agree with earlier calculations employing \citet{1975Ap&SS..36..459K}.
Figure~\ref{fig3} shows the calcium K line in two polluted, magnetic white dwarfs. In the case of NLTT~7547, the broad line shape requires a field spread characteristic of a centered dipole ($a_z=0$) of 240 kG, while in the case of NLTT~53908, the narrow Zeeman components require a marked offset ($a_z=-0.2$) and a dipole field of 635 kG.

\section{Discussion}

\citet{2014MNRAS.439L..90K} found evidence of field enhancement among cool, polluted hydrogen-rich white dwarfs. This simple fact can be interpreted either as evidence of a correlation between magnetic field strength and heavy element pollution, or as a field enhancement in {\it all} cool white dwarfs.

Ultimately, this project aims at delivering field structure and binary properties for a large sample of magnetic white dwarfs and constrain population statistics. In particular we seek to determine the fraction of magnetic white dwarfs as a function of age, companionship, and spectral type.
 
\acknowledgements
A.K., L.F. and S.V. acknowledge support from the Czech Science Foundation (15-15943S). This work is based on observations made with ESO telescopes at the La Silla Paranal Observatory under programme IDs 84.D-0862, 90.D-0473, 091.D-0267 and 095.D-0311, and at Kitt Peak National Observatory and Cerro Tololo Inter-American Observatory (National Optical Astronomy Observatory).

\bibliography{vennes_v2}

\begin{thebibliography}{12}
\expandafter\ifx\csname natexlab\endcsname\relax\def\natexlab#1{#1}\fi

\bibitem[{{Achilleos} \& {Wickramasinghe}(1989)}]{1989ApJ...346..444A}
{Achilleos}, N. \& {Wickramasinghe}, D.~T. 1989, {\it \apj}, {\bf 346}, 444

\bibitem[{{Garstang} \& {Kemic}(1974)}]{1974Ap&SS..31..103G}
{Garstang}, R.~H. \& {Kemic}, S.~B. 1974, {\it \apss}, {\bf 31}, 103

\bibitem[{{Kawka} {et~al.}(2017){Kawka}, {Briggs}, {Vennes}, {Ferrario},
  {Paunzen}, \& {Wickramasinghe}}]{2017MNRAS.466.1127K}
{Kawka}, A., {Briggs}, G.~P., {Vennes}, S., {et~al.} 2017, {\it \mnras}, {\bf
  466}, 1127

\bibitem[{{Kawka} \& {Vennes}(2011)}]{2011A&A...532A...7K}
{Kawka}, A. \& {Vennes}, S. 2011, {\it \aaa}, {\bf 532}, A7

\bibitem[{{Kawka} \& {Vennes}(2012)}]{2012MNRAS.425.1394K}
{Kawka}, A. \& {Vennes}, S. 2012, {\it \mnras}, {\bf 425}, 1394

\bibitem[{{Kawka} \& {Vennes}(2014)}]{2014MNRAS.439L..90K}
{Kawka}, A. \& {Vennes}, S. 2014, {\it \mnras}, {\bf 439}, L90

\bibitem[{{Kawka} {et~al.}(2007){Kawka}, {Vennes}, {Schmidt}, {Wickramasinghe},
  \& {Koch}}]{2007ApJ...654..499K}
{Kawka}, A., {Vennes}, S., {Schmidt}, G.~D., {Wickramasinghe}, D.~T., \&
  {Koch}, R. 2007, {\it \apj}, {\bf 654}, 499

\bibitem[{{Kemic}(1975)}]{1975Ap&SS..36..459K}
{Kemic}, S.~B. 1975, {\it \apss}, {\bf 36}, 459

\bibitem[{{Landi Degl'Innocenti} \& {Landolfi}(2004)}]{2004ASSL..307.....L}
{Landi Degl'Innocenti}, E. \& {Landolfi}, M., eds. 2004, Astrophysics and Space
  Science Library, Vol. 307, {Polarization in Spectral Lines}

\bibitem[{{Landstreet} {et~al.}(2017){Landstreet}, {Bagnulo}, {Valyavin}, \&
  {Valeev}}]{2017A&A...607A..92L}
{Landstreet}, J.~D., {Bagnulo}, S., {Valyavin}, G., \& {Valeev}, A.~F. 2017,
  {\it \aaa}, {\bf 607}, A92

\bibitem[{{Martin} \& {Wickramasinghe}(1984)}]{1984MNRAS.206..407M}
{Martin}, B. \& {Wickramasinghe}, D.~T. 1984, {\it \mnras}, {\bf 206}, 407

\bibitem[{{Zuckerman} {et~al.}(2003){Zuckerman}, {Koester}, {Reid}, \&
  {H{\"u}nsch}}]{2003ApJ...596..477Z}
{Zuckerman}, B., {Koester}, D., {Reid}, I.~N., \& {H{\"u}nsch}, M. 2003, {\it
  \apj}, {\bf 596}, 477

\end{thebibliography}
\end{document}